\documentclass[referee,a4paper,12pt,traditabstract]{swsc}

\usepackage{graphicx}
\usepackage{subfigure}
\usepackage{epstopdf}
\usepackage{lineno}
\usepackage[authoryear,round]{natbib}
\usepackage{hyperref}
\usepackage{url}
\usepackage{rotating}

\hypersetup{colorlinks=true,citecolor=cyan,urlcolor=cyan,linkcolor=blue}

\begin{document}


   \title{Heliospheric tracking of enhanced density structures of the 6 October 2010 CME}
   
   \titlerunning{Heliospheric tracking of density structures of CME}

   \authorrunning{Mishra and Srivastava}

   \author{Wageesh Mishra
          \and
          Nandita Srivastava
          }

   \institute{Udaipur Solar Observatory, Physical Research Laboratory, Udaipur, India\\
              \email{\href{wageesh@prl.res.in}{wageesh@prl.res.in}}
             }
\date{Received --; Accepted --}
 
		\abstract
{A Coronal Mass Ejection (CME) is an inhomogeneous structure consisting of different features which evolve differently with the propagation of the CME. Simultaneous heliospheric tracking of different observed features of a CME can improve our understanding about relative forces acting on them. It also helps to estimate accurately their arrival times at the Earth and identify them in in- situ data. This also enables to find association between remotely observed features and in-situ observations near the Earth. In this paper, we attempt to continuously track two density enhanced features, one at the front and another at the rear edge of the 6 October 2010 CME. This is achieved by using time-elongation maps constructed from \textit{STEREO}/SECCHI observations. We derive the kinematics of the tracked features using various reconstruction methods. The estimated kinematics are used as inputs in the Drag Based Model (DBM) to estimate the arrival time of the tracked features of the CME at L1. On comparing the estimated kinematics as well as the arrival times of the remotely observed features with in-situ observations by \textit{ACE} and \textit{Wind}, we find that the tracked bright feature in the \textit{J}-map at the rear edge of 6 October 2010 CME corresponds most probably to the enhanced density structure after the magnetic cloud detected by \textit{ACE} and \textit{Wind}. In-situ plasma and compositional parameters provide evidence that the rear edge density structure may correspond to \textit{a} filament associated with the CME while the density enhancement at the front corresponds to the leading edge of the CME. Based on this single event study, we discuss the relevance and significance of heliospheric imager (HI) observations in identification of the three-part structure of the CME.}


   \maketitle

\section{Introduction}
\label{Intro}

A CME is huge magnetised plasma eruption from the Sun into the heliosphere and can be observed due to Thomson scattering of photospheric light off free electrons in the corona \citep{Billings1966,Howard2009}. In coronagraphic observations, a classic CME displays the so-called three-part structure. This includes a leading edge which is followed by a dark cavity and a bright core \citep{Illing1985}. The Solar TErrestrial Relations Observatory (\textit{STEREO}) \citep{Howard2008} Heliospheric Imager (HI) \citep{Eyles2009} era has proved to be a boon for a solar terrestrial physicist in tracking a CME. However, the identification of different features of CMEs in the heliosphere by their continuous tracking and the prediction of their arrival time at 1 AU has been achieved with limited accuracy only \citep{Liu2010,Howard2012a,Liu2013,Mishra2013,Mishra2014a,Mishra2014b}. Due to lack of information regarding the evolution of a CME during its propagation between the Sun and the Earth, and the process by which it manifests itself in ambient solar wind, sometimes, its association in in-situ observations is difficult. This is in part because of the challenges in extracting the faint Thomson scattered signal from the brighter background signals dominated by instrumental stray light, F-corona and background starfield \citep{Eyles2009}. The heliospheric tracking of CMEs using \textit{J}-maps \citep{Sheeley1999,Davies2009} mostly deals with tracking of a CME front and associating it with the sheath observed prior to the leading edge in in-situ data \citep{Davis2009,Liu2010,Liu2011,Mostl2011,Mishra2013}. In a rare attempt, \citet{Howard2012a} tracked a cavity like feature (in coronagraph images) of a classical CME using HI images which could be associated with a magnetic cloud identified in in-situ data near the Earth. Also, \citet{Deforest2011} have attempted to identify different CME structures in \textit{STEREO}/HI2 observations and compared them with in-situ features detected near 1 AU. Sometimes, the observed features in coronagraphic images become too faint to be detected in HI images and to be tracked in \textit{J}-maps, especially for an Earth-directed feature when \textit{STEREO} is behind the Sun. In the absence of continuous tracking of a CME from the Sun to the Earth, the extrapolation of obtained heliospheric kinematics out to the Earth or use of these kinematics in appropriate models (e.g. drag based model) is done currently to predict the arrival time of CME at L1 \citep{Liu2010,Colaninno2013,Mishra2013}.

Over the last decade, with the availability of heliospheric imaging observations, a number of reconstruction methods have been developed by various authors \citep{Tappin2004,Howard2006,Kahler2007,Sheeley2008,Lugaz2009,Lugaz2010a,Liu2010,Davies2012,Davies2013}. These methods are based on a number of assumptions on the geometry and evolution of CMEs. Therefore, prediction of CME arrival time, even if tracked continuously out to larger elongations, is not very accurate \citep{Davis2010,Lugaz2010,Mishra2014}. The reconstruction methods enable the estimation of kinematics using the elongation variations of solar wind transients like CMEs. Therefore, the accuracy in remote identification, tracking and extraction of elongation angles of various CME features is of prime concern for their association with in-situ data and for the estimation of accurate arrival time.

A CME shows large scale inhomogeneous structures in terms of density and magnetic field. Therefore, it is likely that these structures will be acted upon by unequal forces which can result in different kinematics. Occasionally, if a CME feature is missed by an in-situ spacecraft, the sequential tracking of other structures might be helpful in relating the remote observations and in-situ. Such a study is of two-fold importance. Scientifically, the understanding of the physical nature of various features (leading edge, cavity, and core) of a CME can help in the theoretical modeling of a CME to investigate its heliospheric evolution. On the other hand, different features of a CME may lead to different perturbations in the Earth's magnetosphere because of their dissimilar plasma and magnetic field parameters, which also needs to be understood.

In this paper, we report on the evolution of the front and the rear edge of a filament associated geo-effective CME of the 6 October 
2010. The layout of paper is as follows: In Section~\ref{Obs}, we present the observations of the CME. In Section~\ref{Analy}, we describe the methodology followed for the analysis of tracking of different features of the CME in COR and HI images. We discuss the results in Section~\ref{Resdis}.

\section{CME of 6 October 2010: Observations and Analysis Approach}
\label{Obs}
\begin{figure}
\begin{center}
\includegraphics[angle=0,scale=1.9]{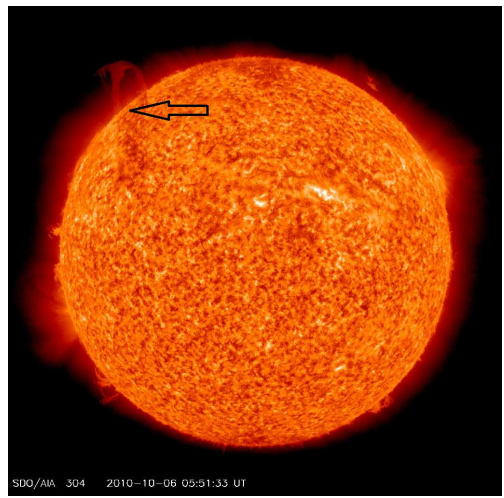}
\caption{The erupting filament material, shown with an arrow, on the NE quadrant of the Sun, observed in 304 {\AA} image from SDO/AIA.}
\label{Fila}
\end{center}
\end{figure}

We use the white light observations of the CME from twin spacecraft \textit{STEREO} mission and in-situ plasma and magnetic field parameters of the solar wind from the Advanced Composition Explorer \textit{ACE} \citep{Stone1998} and \textit{Wind} \citep{Ogilvie1995} spacecraft. \textit{STEREO-A} and \textit{STEREO-B} move in such a way that the separation between them increases by about 45$^\circ$ per year. Each \textit{STEREO} carries an identical imaging suite, called the Sun Earth Connection Coronal and Heliospheric Investigation (SECCHI: \citealp{Howard2008}). SECCHI consists of five telescopes, namely, a EUV imager, two coronagraphs (COR1 and COR2) and two heliospheric imagers (HI1 and HI2). Combining the field-of-view (FOVs) of both coronagraphs and both heliospheric imagers together, a CME can be imaged from elongation 0.4$^\circ$ to 88.7$^\circ$, i.e. from its birth in the inner corona all the way to the Earth and beyond. The \textit{ACE} and \textit{Wind} spacecraft are located at the first Lagrangian point (L1) and measure the plasma and magnetic field parameters of the solar wind. At the launch time of the CME on 6 October 2010, the separation between the twin \textit{STEREO} spacecraft was about 161$^\circ$. The CME was associated with a prominence eruption located in the North-East 
(NE) quadrant of the solar disc as seen from the Earth's perspective. Figure~\ref{Fila} shows the filament eruption in 304 {\AA} obtained from the Solar Dynamic Observatory/Atmospheric Imaging Assembly (SDO/AIA) \citep{Lemen2012} .

\begin{figure}
\begin{center}
\includegraphics[angle=0,scale=.22]{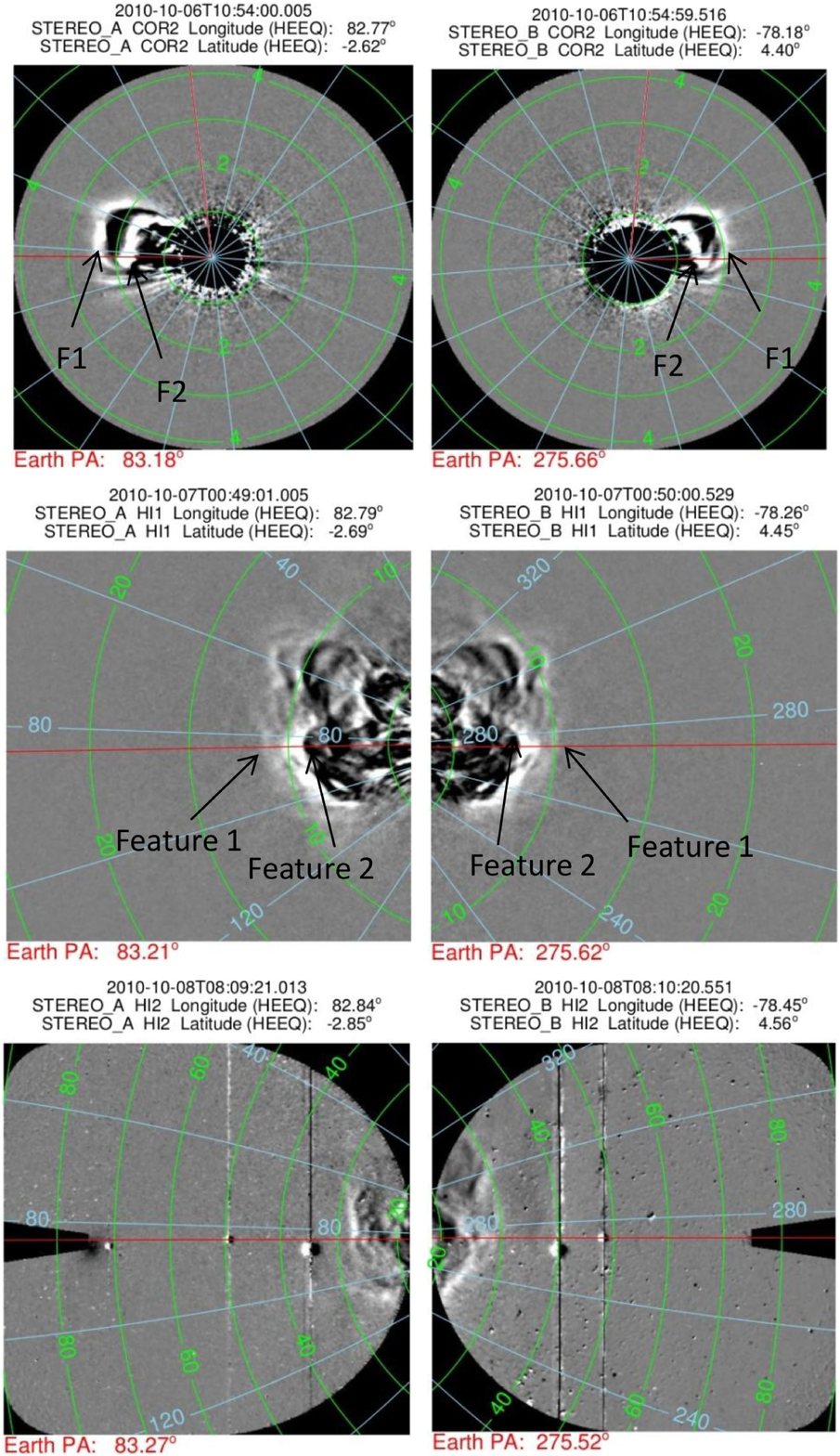}
\caption{Evolution of the 6 October 2010 CME observed in COR2, HI1 and HI2 images from \textit{STEREO-A} (left column) and \textit{STEREO-B} (right column). Contours  of elongation angle (green) and position angle (blue) are overplotted. The vertical red line in the COR2 images marks the 0$^\circ$ position angle contour. The horizontal lines (red) on all panels indicate the position angle of Earth. The features (F1 and F2) tracked in coronagraph FOVs and the features (Feature 1 and Feature 2) tracked in the HI FOV are shown with arrows (black) in the top and middle panels, respectively.}
\label{Evolution}
\end{center}
\end{figure}

The evolution of the 6 October 2010 CME in COR and HI FOV is shown in Figure~\ref{Evolution}.  The front/leading edge (termed F1) and core/associated filament  (termed F2) could be easily identified in COR1 and COR2 images. These features are marked with arrows (in black) in the top panels of Figure~\ref{Evolution}. By manual selection of the features F1 and F2 in both sets of twin \textit{STEREO} images and application of the tie-pointing method of 3D reconstruction, their kinematics are estimated. As the CME approached the edge of the COR2 FOV, \textit{STEREO-A} had a data gap until the CME appeared in the HI1-A FOV as a diffuse structure where its leading edge and core could not be distinguished as they could in the COR FOV. Therefore, we constructed the \textit{J}-maps for this CME using running difference images of HI1 and HI2 data and tracked the two density enhanced features at its front and rear edge. In the  \textit{J}-map the tracked features are termed Feature 1 and Feature 2 as we are not certain that these features are exactly the same as F1 and F2 tracked in the COR FOV. Feature 1 and Feature 2 are each marked with an arrow (in black) in the middle panel of Figure~\ref{Evolution}. Using derived elongation from \textit{J}-maps for these features, we estimated their kinematics. Based on their arrival times, we attempted to identify them in in-situ observations and associate them with the observed features in coronagraph observations. 

In STEREO era, for estimating the kinematics of CMEs, several 3D reconstruction methods have been developed based on different assumptions regarding geometry, speed and propagation direction, Thomson scattering geometry, and the use of observations from single or multiple viewpoints. In a recent study, we have shown that different reconstruction methods (applied to HI observations) estimate different kinematic profiles of a CME \citep{Mishra2014}. We also show that stereoscopic methods which take into account the spherical geometry of a CME are superior to single spacecraft reconstruction methods and single  spacecraft fitting methods. But to take into account the maximum uncertainties in our analysis, we use a total of 10 reconstruction methods. Use of these methods also allow us to verify their performance by estimating the kinematic profiles and arrival time  of tracked features. We have implemented seven single spacecraft methods, Point-P (PP: \citealp{Howard2006}), Fixed-Phi (FP: \citealp{Kahler2007}), Harmonic Mean (HM: \citealp{Lugaz2009}), Self-similar expansion (SSE: \citealp{Davies2012}), Fixed-Phi Fitting (FPF: \citealp{Rouillard2008}), Harmonic Mean Fitting (HMF: \citealp{Lugaz2010}) and Self- Similar Expansion Fitting (SSEF: \citealp{Davies2012}), which require elongation measurements of a moving CME feature from a single viewpoint. We have also implemented three stereoscopic methods, namely  Geometric Triangulation (GT: \citealp{Liu2010a}), Tangent to A Sphere (TAS: \citealp{Lugaz2010a}) and Stereoscopic Self-Similar Expansion (SSSE: \citealp{Davies2013}), which require simultaneous elongation measurements from two viewpoints to estimate the kinematics of a tracked feature. It is to be noted that in FP, GT and FPF reconstruction methods, a CME is treated as a point, while PP, HM, HMF, SSE, SSEF, TAS, and SSSE reconstruction methods consider a CME as a large-scale structure. In FPF, HMF and SSEF methods, a CME is assumed to propagate with a constant speed along a fixed direction.  Therefore, different assumptions make these methods independent of each other to some extent \citep{Mishra2014}.

We used the estimated kinematics as inputs in the \citet{Vrsnak2013} drag based model (DBM), for the distance over which the CME could not be tracked unambiguously in \textit{J}-maps, and estimated its arrival time at L1. The  \citet{Vrsnak2013} DBM model is based on the assumption that after a distance of 20 R$_\odot$, the dynamics of a CME is governed completely by drag forces \citep{Cargill1996,Cargill2004} acting on it and the drag acceleration is proportional to quadratic of the difference in speed of the CME and ambient solar wind. The proportionality constant is called the drag parameter, which lies in the range from 0.2 $\times$ 10$^{-7}$ to 2.0 $\times$ 10$^{-7}$ km$^{-1}$ as per their analysis. We used these values of drag parameter and chose the speed of the ambient solar wind medium as 350 km s$^{-1}$, in agreement with the prevailing slow ambient solar wind speed for this CME. The obtained kinematics and arrival times of both tracked features are used to associate them with density enhanced structures observed in in-situ observations. We also attempt to find an association between the three-part structure of the CME seen in the COR FOV with features observed in HI and in in-situ data.

\section{Heliospheric Tracking of the 6 October 2010 CME features}
\label{Analy}
\subsection{Remote Sensing Observations}
\label{Rmtana}
\begin{figure}
\begin{center}
\includegraphics[angle=0,scale=1.0]{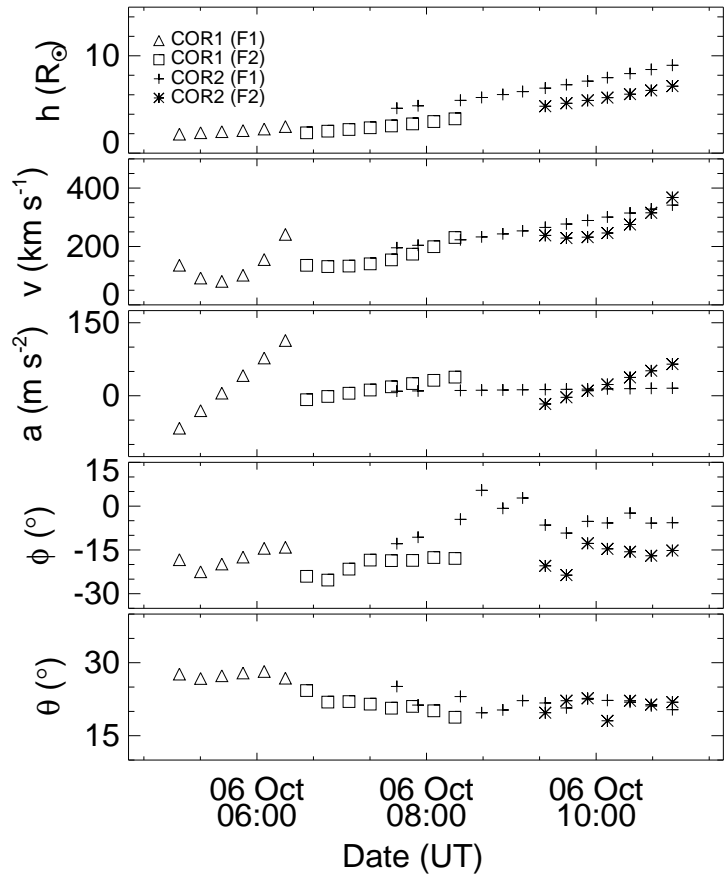}
\caption{Top to bottom panels show the evolution of height, speed, acceleration, longitude and latitude. Here F1 and 
F2 correspond to features on the leading edge and core/filament of the 6 October 2010 CME.}
\label{3DCOR}
\end{center}
\end{figure}

In both sets of SECCHI/COR images, the front (leading edge) and filament (rear edge) of the 6 October 2010 CME are identified clearly. We applied the tie-pointing method of 3D reconstruction (scc\_measure: Thompson 2009) to feature on the leading edge (F1) and in the filament/core (F2) of the CME observed in COR1 and COR2 images to estimate their 3D dynamics. Prior to the implementation of the 3D reconstruction technique, coronagraphic images were processed following the procedure as described by \citet{Mierla2009}. The data from the remote sensing instruments on STEREO is taken from UK Solar System Data Centre (http://www.ukssdc.ac.uk/). The estimated 3D location, in the stonyhurst heliographic coordinate system, for tracked features (F1 \& F2) are plotted in Figure~\ref{3DCOR}. We admit the possibility of human error in the identification of features in the pair of STEREO images on which tie-pointing is performed. An earlier study by \citet{Joshi2011} shows that an error of three pixels leads to errors of 0.12 \textit{R}$_\odot$ and 0.6 \textit{R}$_\odot$ in the distance estimates in the COR1 and COR2 FOVs, respectively while the errors in longitude and latitude are less than 2$^\circ$. Such negligibly small errors have no crucial implications on the results obtained in the present study.

From Figure~\ref{3DCOR}, it is clear that both F1 \& F2 move at around 30$^\circ$ to 20$^\circ$ north from the ecliptic in the coronagraphic FOV and are eastward of the Sun-Earth line. The estimated kinematics show that they follow approximately the same trajectory in 3D space. The leading edge (F1) shows acceleration in the COR1 FOV while filament (F2) seems to accelerate more in the COR2 FOV than in the COR1 FOV.  The leading edge (F1) becomes too diffuse to be tracked at the time when filament (F2) reached the COR1 FOV. From Figure~\ref{3DCOR}, it is clear that the separation between the features F1 and F2 is approximately 1.0 R$_\odot$ in the COR1 FOV. F1 and F2 could also be tracked in the COR2 FOV up to 10:54 UT on 6 October 2010. Because of a data gap from 10:54 UT until 17:39 UT in science images from \textit{STEREO-A}, these features could not be tracked and reconstructed further. At the outermost tracked points in the COR2 FOV, the separation between F1 and F2 increased to 3 R$_\odot$.

\begin{figure}
\begin{center}
\includegraphics[height=9cm, width=8cm]{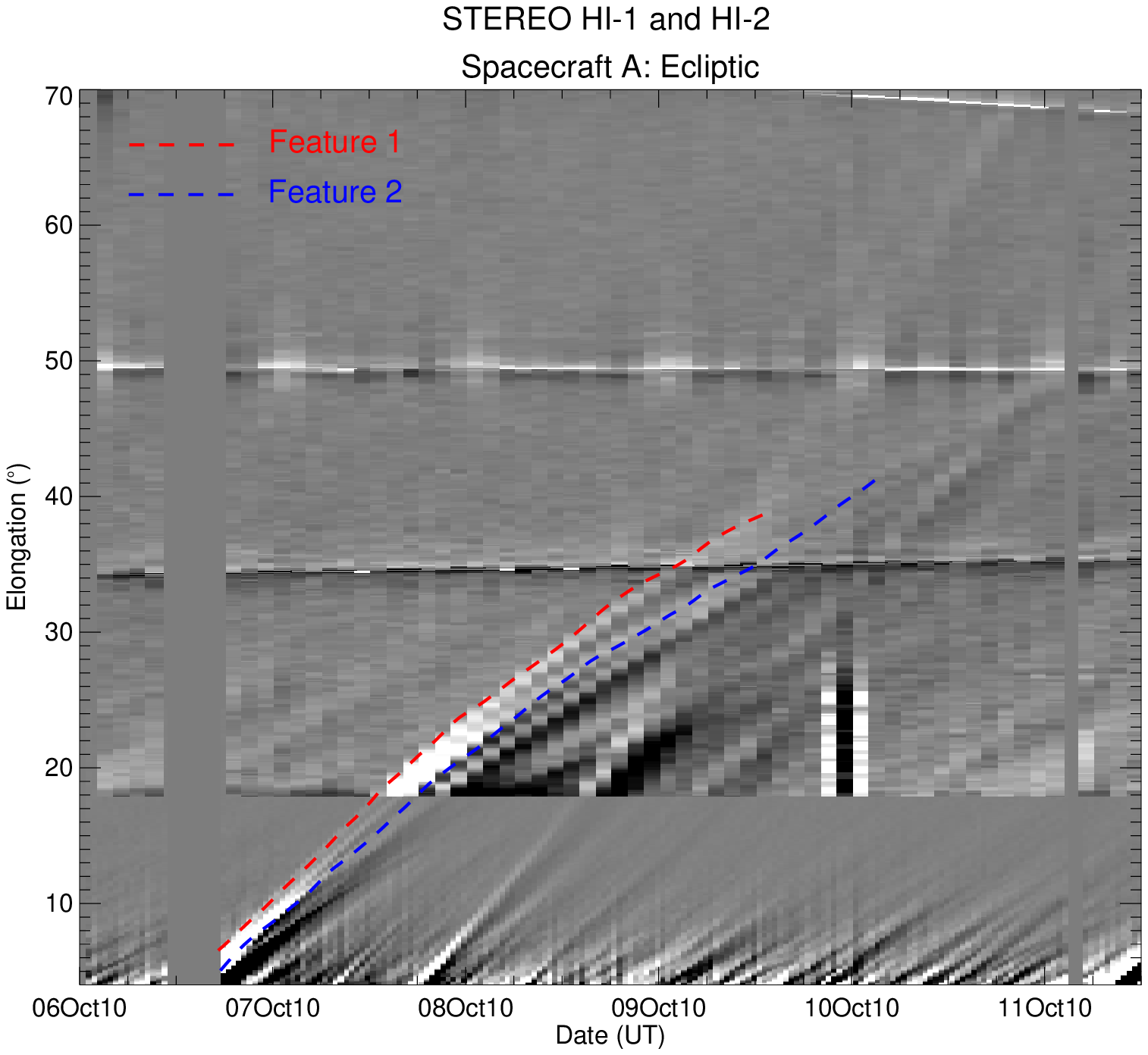}
\hspace{0.3cm}
\includegraphics[height=9cm, width=8cm]{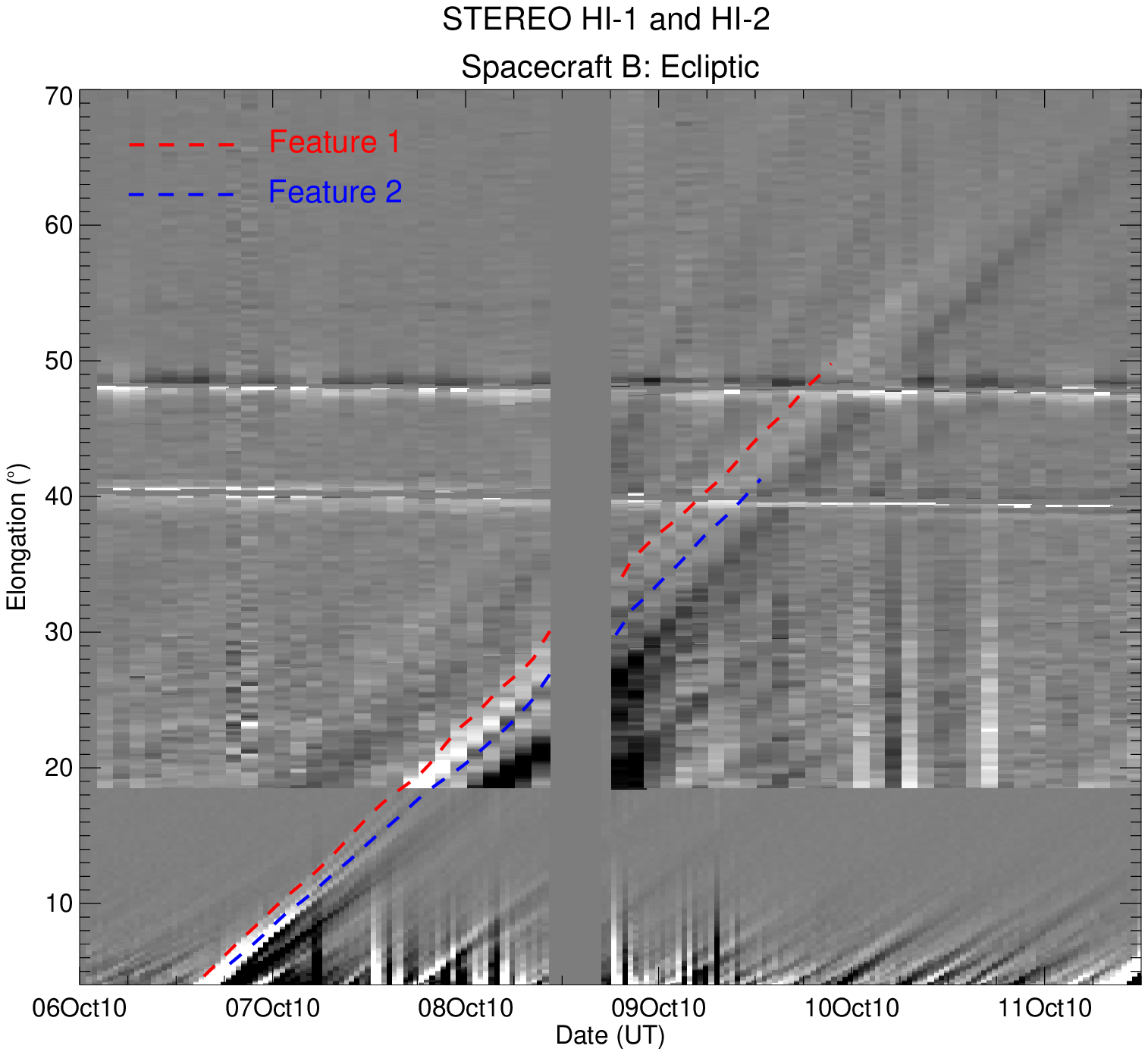}
\caption{Time-elongation maps (\textit{J}-maps) for \textit{STEREO-A} (left) and \textit{STEREO-B} (right) constructed from running differences images from HI1 and HI2, for the time interval from 06 Oct 2010 to 11 Oct 12:00 UT, 2010. Two features (marked as Feature 1 and Feature 2, with red and blue lines, respectively) are tracked corresponding to enhanced density features.}
\label{Jmaps}
\end{center}
\end{figure}

To track the CME features in the HI FOV, we used the \textit{J}-map technique \citep{Rouillard2008,Sheeley2008,Davies2009,Harrison2012} based on a method originally developed by \citet{Sheeley1999} to track features in coronagraph FOVs. We constructed the \textit{J}-maps using running difference images of HI1 and HI2. The detailed procedure is explained in \citet{Mishra2013}. Figure~\ref{Jmaps} shows \textit{J}-maps  where two positively inclined bright tracks which correspond to outward motion of two density structures can be seen. These bright tracks (marked with red and blue)  could be due to features of two different CMEs or different features (front and rear edge) of the same CME passing along the ecliptic. We examined the background subtracted movies from COR and HI, COR1 CME catalogue (\url{http://cor1.gsfc.nasa.gov/catalog/}) and the LASCO CDAW catalogue (\url{http://cdaw.gsfc.nasa.gov/CME_list/}), and found that no Earth-directed CME was launched close to the occurrence of the 6 October 2010 CME. Therefore, we assume that these tracked features are two different features of the Earth-directed geo-effective CME of 6 October 2010. The presence of the  planets Venus and Earth at elongations of 34.3$^\circ$ and 49.5$^\circ$, respectively, in the HI2-A FOV caused the two horizontal lines in the left panel of Figure~\ref{Jmaps}. In the right panel of this figure, two horizontal lines are due to presence of these planets at elongations of 40.4$^\circ$ and 48.2$^\circ$ in the HI2-B FOV. A slanted line that appears in the left panel of Figure~\ref{Jmaps} on 10 October is due to the entrance of Jupiter into the HI2-A FOV.

We derived the elongation variation of both moving features (Feature 1 and Feature 2) by tracking the density enhancements that appear as positively inclined bright tracks in the \textit{J}-maps (Figure~\ref{Jmaps}). These elongation angles are converted to distance using various reconstruction methods described in Section~\ref{Obs}. Feature 1 and Feature 2 could be tracked out to 39$^\circ$ and 41$^\circ$ elongation, respectively, in \textit{STEREO-A} \textit{J}-maps. In \textit{STEREO-B} \textit{J}-maps, Features 1 and 2 could be tracked out to 50$^\circ$ and 41$^\circ$, respectively. In our earlier paper \citep{Mishra2014}, the analysis of the front edge (Feature 1) of the 6 October 2010 CME is carried out by estimating its kinematics using all 10 methods (PP, FP, FPF, HM, HMF, SSE, SSEF, GT, TAS, SSSE) and their results are compared in details. The obtained kinematics are used as inputs to the \citet{Vrsnak2013} DBM to estimate the arrival time (for details see, Figure 3, 4, 5 and Table 1 in \citealp{Mishra2014}). In the current paper, we focus on the analysis of the second feature (Feature 2).

\begin{figure}
\begin{center}
\includegraphics[angle=0,scale=0.7]{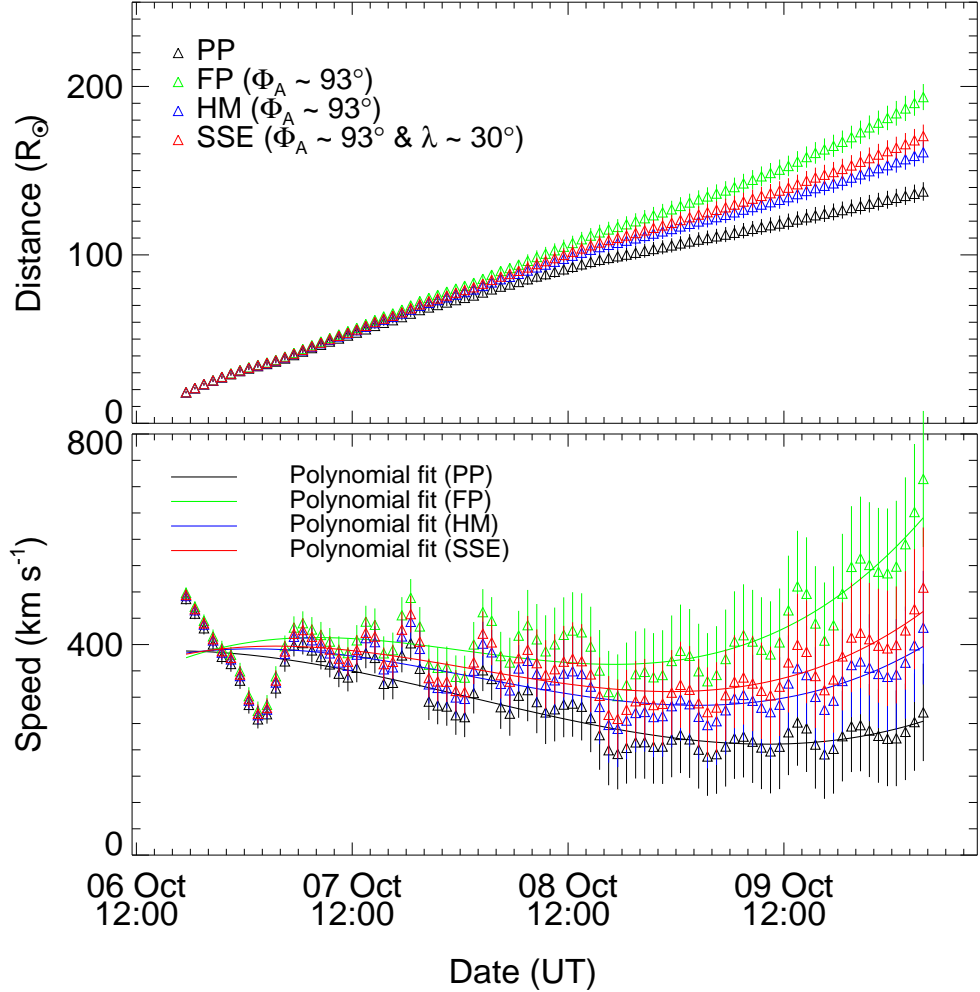}
\caption{Top panel shows the estimated distance profiles of Feature 2, based on application of the PP, FP, HM and SSE methods based on elongation-time variations estimated from \textit{STEREO-A} \textit{J-}maps. In the bottom panel, speed profiles derived from the adjacent distances using three point Lagrange interpolation (solid line shows the polynomial fit) are shown. Vertical lines show the error bars, derived by taking uncertainties of 3\% and 4\% in the estimated distance in the  HI1 and HI2 FOV, respectively.}
\label{KinAA}
\end{center}
\end{figure}

\begin{figure}
\begin{center}
\includegraphics[angle=0,scale=0.7]{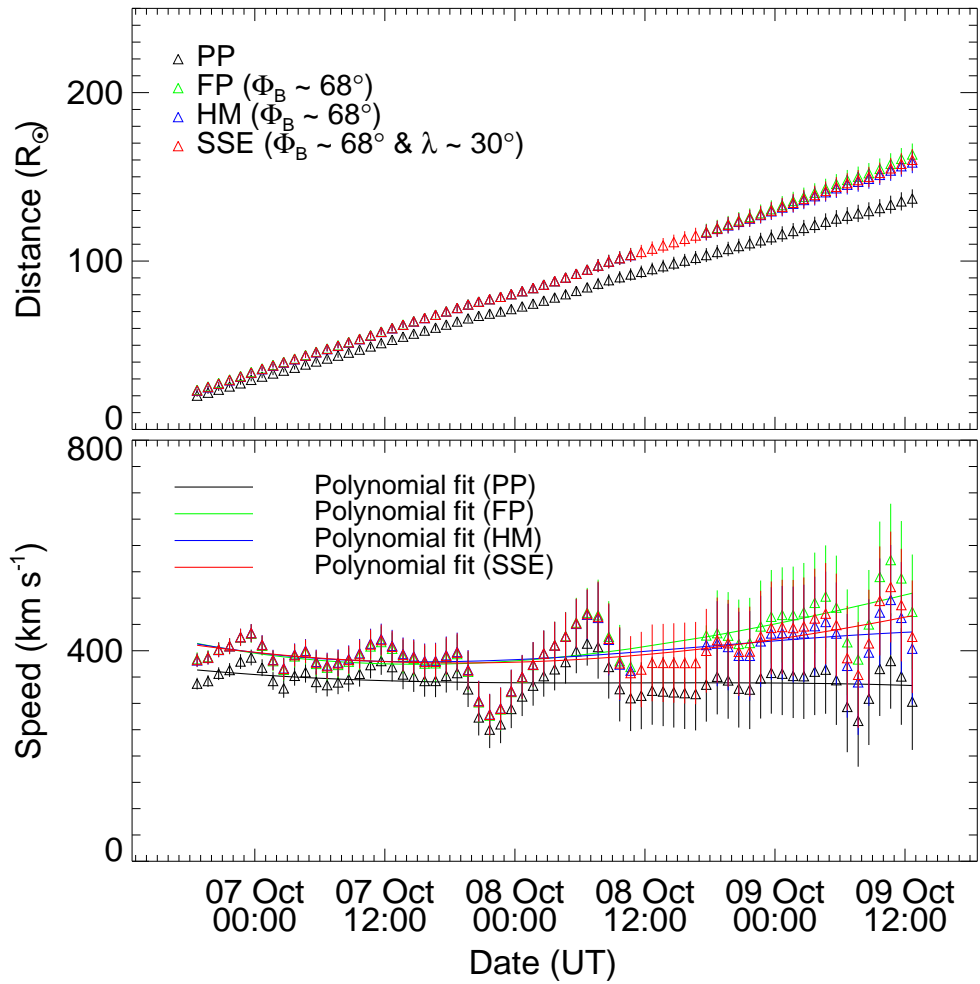}
\caption{As Figure~\ref{KinAA}, corresponding to application of methods on elongation-time variations estimated from 
\textit{STEREO-B} \textit{J-}maps.}
\label{KinBB}
\end{center}
\end{figure}

We used the single spacecraft PP, FP, HM and SSE methods to derive the kinematics of the tracked Feature 2. The FP, HM and SSE methods require the direction of propagation (longitude) of the tracked feature, which is obtained from the longitude (Figure~\ref{3DCOR}) estimated using the tie-pointing reconstruction method in the COR2 FOV. We fixed the propagation direction of this feature as 10$^\circ$ East of the Sun-Earth line, which corresponds to a longitude difference of 93$^\circ$ and 68$^\circ$ from \textit{STEREO-A} and \textit{B}, respectively. In SSE method, we need to fix the angular half-width ($\lambda$) of the CME \citep{Davies2012}  which is taken as 30$^\circ$ in our case. The parameter $\lambda$ is related to the curvature of the CME front and hence it measures the width of the CME. The kinematics profile of tracked Feature 2 from \textit{STEREO-A} and \textit{STEREO-B} viewpoints is estimated and shown in Figure~\ref{KinAA} and Figure~\ref{KinBB}, respectively. The estimated kinematics over the outermost tracked points have been used in the \citet{Vrsnak2013} DBM corresponding to extreme range of drag parameter to obtain its arrival time at L1 (Table~\ref{KinarrL1}). We also applied the three fitting methods (FPF, HMF and SSEF) to obtain a set of speed, propagation direction and launch time parameters that best reproduce the observed elongation-time profiles (extracted from \textit{STEREO-A} and \textit{B} \textit{J}-maps) of Feature 2. The angular half-width of the CME is taken to be 30$^\circ$ when applying the SSEF method. The retrieved best-fit parameters are used to estimate the arrival time of Feature 2 at L1 and are given in Table~\ref{KinarrL1}.

\begin{figure}
\begin{center}
\includegraphics[angle=0,scale=0.7]{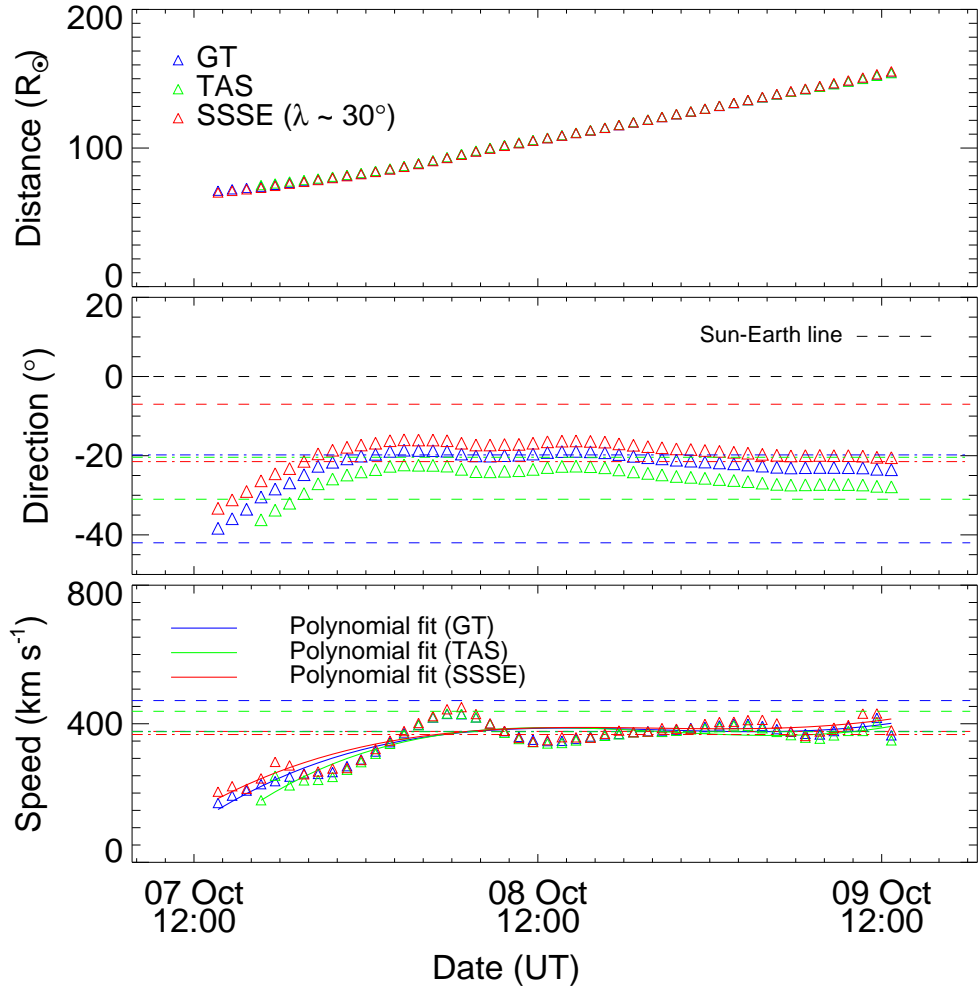}
\caption{From top to bottom, panels show the estimated distance, propagation direction and speed of the tracked Feature 2 using the GT, TAS and SSSE methods. In the middle and bottom panels, the direction and speed estimates from the single-spacecraft fitting methods are also overplotted. In these panels, dashed and dashed-dotted lines correspond to estimates from \textit{STEREO-A} and \textit{STEREO-B} observations, respectively. For these lines, the colors red, blue and green correspond to FPF, HMF and SSEF methods. In the middle panel, dashed horizontal line (black) marks the Sun-Earth line.}
\label{KinAABB}
\end{center}
\end{figure}

For associating remotely observed features with in-situ observations, we must correctly ascertain their direction of propagation in the heliosphere. The direction of propagation of tracked Feature 1 has been estimated using various methods and was found to be toward the Earth \citep{Mishra2014}. If Feature 2 is not following the direction of propagation of Feature 1, then the same elongation for both the features will correspond to a different distance from the Sun. Therefore, we must examine the continuous evolution of Feature 2, the direction of propagation of which  can not be determined using single spacecraft methods (e.g. PP, FP, HM, SSE, FPF, HMF, SSEF). For this, we also implemented the stereoscopic reconstruction methods, i.e. GT, TAS and SSSE (with $\lambda$ = 30$^\circ$) to estimate the kinematics of Feature 2 in the heliosphere. The GT and TAS methods are the triangulation version of the FP and HM methods. The GT and TAS methods assume extreme CME geometries while in the SSSE method the spherical structure for a CME can be fixed to an arbitrary size. \citet{Davies2013} have shown that the TAS and GT methods are special cases of the SSSE method. The details of these methods and their application to Feature 1 are discussed in \citet{Mishra2014}. The obtained kinematics of Feature 2 using the GT, TAS and SSSE methods are presented in Figure~\ref{KinAABB}. In this figure, the kinematics at the sun-ward edge of the HI1 FOV is not shown due to the occurrence of a singularity, where small uncertainties in elongation measurements lead to larger errors in kinematics (for details see, \citealp{Liu2010,Mishra2013,Mishra2014,Davies2013}). In the middle and bottom panels of this figure, the direction and speed estimates from the single-spacecraft fitting methods are also overplotted. These estimates from \textit{STEREO-A} and \textit{STEREO-B} observations are plotted with dashed and dashed-dotted lines, respectively. Red, blue and green colors correspond to the FPF, HMF and SSEF methods, respectively.

Further, we input the kinematics estimated over the outer points into the DBM with an ambient solar speed of 350 km s$^{-1}$ and the extremes of the range of the drag parameter. The estimated arrival time, transit speed at L1 and errors therein are given in Table~\ref{KinarrL1}. From the fifth and sixth columns of the table, it is clear that the errors in the predicted arrival time and speed are less when the maximum range of drag parameter is used in the DBM, for all the methods except for the three fitting methods. This highlights that a following feature (like Feature 2) of a CME probably experiences a larger drag force during its propagation from the Sun to the Earth. From Table~\ref{KinarrL1}, we also notice that different fitting methods give different estimates of Feature 2's propagation direction. Therefore, in light of our earlier study in \citet{Mishra2014} which shows that stereoscopic methods provide reliable estimates of time variations of CME propagation direction, we rely on the CME direction estimated from GT, TAS and SSSE methods.

From Figure~\ref{KinAA} and ~\ref{KinBB}, we see that the PP and FP methods give the lowest and the highest estimate of distance of the tracked feature, respectively. The estimated distances from HM and SSE methods are intermediate  between those obtained from PP and FP methods. In Figure~\ref{KinAA}, the observed unphysical late acceleration of Feature 2 is attributed to a possible deflection of the tracked feature far from the Sun. Although, the real deflection of features far from the Sun is rarely observed, this effect could result from it not being possible to track the same part of the CME leading edge in each successive image. This is especially true at large elongations where expansion of the overall CME plays an important role, leading to its artificial deflection \citep{Howard2011}. Such effect will be more pronounced for the methods (e.g. FP, GT and FPF) in which the finite size of the CME is not taken into account. The performance of these single-spacecraft methods at larger elongations is sensitive to changes in the input values of propagation direction of CMEs used in the methods. Figure~\ref{KinAA} and ~\ref{KinBB} also highlight that different values of CME propagation direction, used as inputs to the single-spacecraft methods, lead to different values of kinematic parameters. In our analysis, we have taken care that if, at any instant, the estimated kinematics become unreliable, due to breakdown of some assumptions made in the reconstruction methods, then the kinematics estimated prior to those points are used as inputs to the DBM. It  must also be noted that if the position of the Earth or in-situ spacecraft at L1 is at different longitude than that of the ICME apex (the point of the CME leading edge at the largest heliospheric distance from the Sun), then a CME with a circular geometry (as modeled in HM and SSE methods) will lead to delay in the arrival times and lower speeds at the Earth than that based on the CME apex kinematics. In this scenario, a CME can miss to hit the Earth. Therefore, values of speeds must be deprojected along the Sun-Earth line before extrapolating or using them in the DBM for arrival time prediction at L1. The estimated direction of propagation of Feature 2 from implementation of the fitting methods (FPF, HMF, and SSEF) is significantly away (up to 42$^\circ$ east) from the Sun-Earth line, therefore an off-apex correction is applied on the speed obtained from these methods \citep{Mostl2011,Mostl2013}. In principle, such an off-axis correction should also be applied for HM, SSE, TAS, and SSSE methods. However the estimated direction of propagation of Feature 2 from these methods is within $\pm$ 25$^\circ$ from the Sun-Earth line, hence  such a correction would decrease the speed by only a few tens of km s$^{-1}$.

On comparing the two features, Feature 1 (Figure 5 in \citealt{Mishra2014}) and Feature 2 (Figure~\ref{KinAABB}, this paper), we see that both moved approximately in the same direction through the heliosphere. Thus, both these Earth-directed features are likely to be detected by an in-situ spacecraft located at L1. Since, we have no information on the temperature, composition of plasma and magnetic field parameters associated with these tracked features; we can only rely on their estimated arrival time at L1 to relate features in in-situ data with Feature 1 and Feature 2. It is noted that while applying several reconstruction methods, we have considered the same angular width for the leading edge and core of the CME which may not be true. Moreover, we have assumed a convex geometry for both the leading edge and core of the CME. Although, we are aware that the convex geometry may not hold good for the core of the CME. This assumption may be invalid even for leading edge of the CME at larger elongations due to its possible flattening. However, in the absence of knowledge about the geometry and the angular width of the core and the leading edge of the CME, we have used the existing methods based on idealistic assumptions \citep{Davies2013,Mishra2014}, which is a pragmatic approach.

\subsection{Comparison of Kinematics of Tracked Features 1 and 2}
\label{SepF1F2}
\begin{figure}
\begin{center}
\includegraphics[angle=0,scale=0.8]{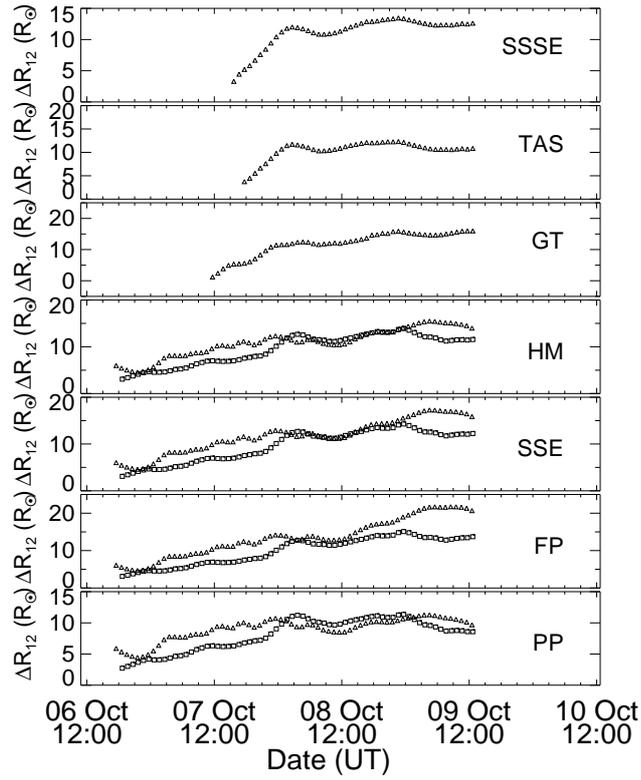}
\caption{Variation in separation distance between the two tracked features with time. Each panel shows computed separation of the two features, based on the use of different methods for the estimation of distance of tracked Feature 1 and Feature 2. From top to bottom, panels correspond to SSSE, TAS, GT (twin spacecraft reconstruction), HM, SSE, FP, and PP (single-spacecraft reconstruction) methods. In the four panels from bottom to top, the results derived by implementing the methods on \textit{STEREO-A} and \textit{STEREO-B} observations are shown with triangles and squares, respectively.}
\label{CompF1F2}
\end{center}
\end{figure}

We tracked the two different features as the front (leading) and rear (trailing) edge of the CME in the \textit{STEREO}/HI \textit{J}-maps and derived their kinematics. The estimated separation between Feature 1 and Feature 2 based on their height derived from different reconstruction methods is shown in Figure~\ref{CompF1F2}. We found that the separation of the two features increased with time (Figures~\ref{3DCOR} \& ~\ref{CompF1F2}) from 1.2 R$_\odot$ to 15 R$_\odot$ during their evolution from the COR to HI FOV. The estimated separation may be due to unequal driving forces on different features or due to a overall CME expansion or due to a combination of both. Figure~\ref{CompF1F2} (the four panels from the bottom) shows the separation estimates by using single-spacecraft methods independently on \textit{STEREO-A} and \textit{B} observations. It is evident that the separation increased from around 3 R$_\odot$ to 11.3 R$_\odot$ for PP, to 21.7 R$_\odot$ for FP and to 15.4 R$_\odot$ for HM. For the SSSE, TAS and GT techniques (the three top panels in Figure~\ref{CompF1F2}), the separation between Feature 1 and Feature 2 increased out to 16.0, 12.5 and 13.5 R$_\odot$, respectively and became approximately constant at a distance of $\approx$ 120 R$_\odot$.

Due to a singularity at the sunward edge of the HI1 FOV in GT, TAS and SSSE methods, the separation between the tracked features (Feature 1 and Feature 2) is estimated to be slightly smaller than the observed separation in the COR FOV. However, these initial erroneous estimates of separation can be avoided in the light of results obtained from GT, TAS and SSSE methods at larger elongations and estimates from other methods. Considering all the reconstruction methods applied to the SECCHI/HI observations, the estimated separation is probably some where between 11.2 R$_\odot$ to 21.7 R$_\odot$.

Several authors have shown that, at large distances from the Sun, the kinematic evolution of CMEs is mostly attributable to the drag force between the CMEs and the ambient solar wind \citep{Cargill2004,Manoharan2006,Vrsnak2010}. In the present case, both of the features are tracked out to sufficiently large distances beyond the HI1 FOV where they attain speed close to that of the ambient solar wind, such that the driving forces on Feature 1 and Feature 2 become equal. The expansion speed of a CME at larger distance is more likely due to the high internal thermal pressure of the CME than the ambient solar wind pressure \citep{Schwenn2005,Gopalswamy2009,Michalek2009,Poomvises2010}. In our case, we notice that the expansion speed of the CME is $\approx$ 165 km s$^{-1}$ at the entrance of the HI FOV which becomes negligibly small compared to radial speed of the CME at a distance of 115 R$_\odot$. Therefore, under the aforementioned constraints regarding expansion and equal drag on both features of the CME, it is expected that the final estimate of the separation between Feature 1 and Feature 2 will be approximately maintained out to the L1 point. If a constant speed of 350 km s$^{-1}$ is assumed for both features beyond the last observation points in the HI FOV, then the difference in their arrival time at L1 will range from 6.1 to 11.9 hr.

\subsection{Identification of Remotely Observed Tracked Features in Near-Earth In-Situ Observations}
\label{insituobs}

The formation and evolution of the three part structure of a CME is not well understood. It is agreed that the leading edge which appears bright, is due to the sweeping up of coronal plasma by erupting flux ropes or the presence of pre-existing material in the overlying fields \citep{Riley2008}. This leading edge is identified near the Earth in in-situ observations as the CME sheath region in the solar wind \citep{Forsyth2006}. The darker cavity region is assumed to correspond to a flux rope structure, and is identified as a magnetic cloud (MC) \citep{Klein1982,Burlaga1991} in in-situ observations and as a void in HI observations \citep{Howard2012a}. The inner-most bright feature (CME core) is associated with cold and dense filament material \citep{Webb1987}. However, identification of filament material in the in-situ observations near the Earth is rare, and its heliospheric evolution is not well understood \citep{Crooker2006, Lepri2010}. Also, due to the large distance gap and the difficulty in determining the true evolution of remotely-sensed features, association between two sets of observations (remote and in-situ) are still challenging. 

\begin{figure}
\begin{center}
\includegraphics[angle=0,scale=0.8]{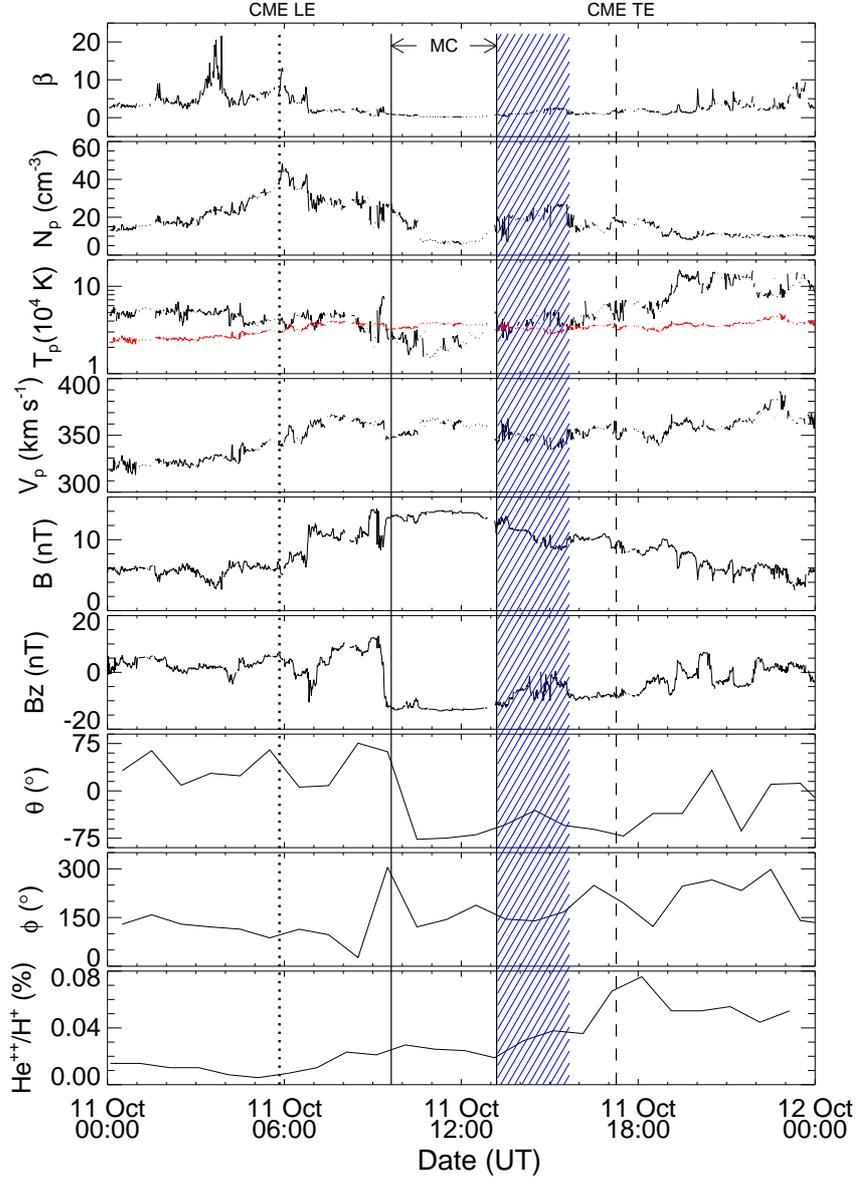}
\caption{From top to bottom, panels show plasma beta, proton density, proton temperature, flow speed, magnitude of magnetic field, z-component of magnetic field, latitude, longitude of magnetic field vector, and alpha to proton ratio. The red curve in the third panel shows the expected proton temperature. From the left, the first 
(LE), second, third and fourth (TE) vertical lines mark the arrival of CME leading edge, start of magnetic cloud, end of magnetic cloud, and trailing edge of CME respectively. The blue hatched regions mark the region associated with tracked Feature 2.}
\label{insitu}
\end{center}
\end{figure}

The in-situ data taken by the \textit{ACE} \citep{Stone1998} and \textit{Wind} \citep{Ogilvie1995} spacecraft are  analysed to identify the CME structures based on plasma, magnetic field and compositional signatures \citep{Zurbuchen2006}. The in-situ observations from 11 to 12 October 2010 are shown in Figure~\ref{insitu}. In this figure, the red curve in the third panel from the top shows the variation of the expected proton temperature described in \citet{Lopez1987} and the first vertical line (dotted, labeled as LE) marks the arrival of the CME leading edge at 05:50 UT on 11 October 2010. The fourth vertical line (dashed, labeled TE) marks the trailing (rear) edge arrival at 17:16 UT. The region bordered by second and third vertical lines (solid), at 09:38 UT and 13:12 UT respectively, can be classified as a MC \citep{Klein1982,Lepping1990}. The enhanced density before the first dotted vertical line is the CME sheath region. \citet{Mishra2014} tracked Feature 1, the leading edge of the initial density front of the 6 October 2010 CME, in \textit{J}-maps and estimated its kinematics based on several methods. The arrival time of Feature 1 is obtained using its kinematics as input to the DBM. Since Feature 1 is the first enhanced density associated with the CME, it is associated with the in-situ measured density enhancement (LE) that arrived at 05:50 UT on 11 October 2010. The time of this density enhancement is in agreement (with some errors) with the predicted arrival time of the CME at L1 \citep{Mishra2014}.

In Figure~\ref{insitu}, we do not notice any monotonic decrease in the speed of the CME, i.e. there is no expansion of MC. This is consistent with the finding that the separation between Feature 1 and Feature 2 became constant well before they reached the L1 point. As discussed in Section~\ref{SepF1F2}, we expect Feature 2 to arrive at L1, approximately 6 to 12 hours after the arrival of Feature 1. In in-situ observations, we found a second peak in proton density between 13:14 and 15:40 UT on 11 October 2010 and, therefore, associated this with arrival of Feature 2. Feature 2 follows the MC, and is associated with dense (maximum N$_{p}$ = 27 cm$^{-3}$), cold material (minimum T$_{p}$ = 2.4 $\times$ 10$^{4}$ K) observed in-situ. It possibly corresponds to the core of a classical three-part structure of a CME. We need to examine the plasma composition and charge state, and the root mean square deviations of bulk velocity to confirm the presence of filament material in the associated CME at L1 \citep{Burlaga1998,Gopalswamy1998,Lepri2010,Sharma2012,Sharma2013}.

Hourly-resolution \textit{ACE}/Solar Wind Ion Composition Spectrometer (SWICS) \citep{Gloeckler1998} data show that the alpha to proton ratio (Figure~\ref{insitu}, lowest panel) is elevated over the entire CME interval peaking at 0.07 (at 18:30 UT on 11 October 2010) which is seven times the ambient solar wind value (before the leading edge of CME). During the second density enhancement between 13:14 and 15:40 UT on 11 October 2010, the alpha to proton ratio is approximately three times greater than its value in the ambient solar wind. The Fe/O ratio is enhanced up to 0.24, which is two times larger than its ambient solar wind value, and also the C$^{+6}$/C$^{+5}$ ratio reduces to 0.5 and its average value is $\approx$ 1.0 thereafter. This enhanced density region corresponds to a decrease in thermal speeds of He$^{+2}$, C$^{+5}$, O$^{+7}$, Fe$^{+10}$  with minimum values of 18, 13, 12 and 13 km s$^{-1}$, respectively. These signatures are indicative of filament plasma passage during 13:14 to 15:40 UT \citep{Burlaga1998,Gopalswamy1998,Sharma2012}. However, during this density  enhancement, the O$^{+7}$/O$^{+6}$ ratio, the average charges state of C (+5), O (+6) and Fe (+10) remain roughly constant although they are expected to decrease in a filament region \citep{Lepri2010}. Therefore, although the CME of 6 October 2010 was associated with an eruptive filament, all the criteria required to definitively identify filament plasma are not fulfilled. This may be due to in-situ observations taken at a single point or may occur due to the heating of the cold filament material during its journey to L1 \citep{Skoug1999,Sharma2012}.

\section{Results and Discussion}
\label{Resdis}
We have attempted to associate two different features of the 6 October 2010 CME in remote and in-situ observations. The two density enhanced features namely Feature 1 and Feature 2 are tracked throughout the combined HI1 to HI2 FOV, and F1 and F2 are tracked from COR1 to COR2. We applied 10 different reconstruction methods to estimate the 3D kinematics of both features. Interestingly, we found that both features were directed towards the Earth and the separation between them increased with time as they evolved in the heliosphere. Our analysis suggests that both these tracked features are associated with density enhancement at the front and the rear edge of the 6 October 2010 CME. The increasing separation between both features (up to 11.2 R$_\odot$-21.7 R$_\odot$) can be attributed to the expansion of the CME or unequal forces acting on them.

Based on the in-situ observations, we marked the observed arrival time of features. These marked arrival times at L1 can be compared with the predicted values at L1 obtained from combining remote-sensing kinematics with DBM. The estimated arrival times, transit speeds and errors therein for the Feature 1 are given in Table 1 of \citet{Mishra2014}. For the tracked Feature 2, estimated arrival time and transit speed at L1 is given in Table~\ref{KinarrL1} of this paper. Also differences in the estimated values from those measured in-situ are listed in our Table~\ref{KinarrL1}. From this table, it is clear that the stereoscopic reconstruction methods which use the simultaneous observation from multiple vantage points give better estimates of the arrival time than those which use observations from single vantage point. This finding is consistent with \citet{Mishra2014} where a comparison of various methods implemented on few CMEs having different characteristics and launched in different solar wind medium was made.

It is found that Feature 1 is associated with the density enhancement in the sheath region of the CME while Feature 2 is associated with the density enhancement at the rear edge of a magnetic cloud. The plasma parameters in the second density structure reveal signatures of filament plasma. Association between remotely tracked Feature 2 and second density enhancement measured in-situ is made relying on the kinematics of both features. Since, filaments follow the cavity (flux rope) in imaging (COR) observations, therefore, based on the in-situ measured plasma and compositional data, we expect that the second peak in density corresponding to Feature 2 at the rear edge of magnetic cloud in in-situ measurements is due to arrival of filament structure. Based on our approach of continuous tracking of density structures at the front and rear edge of CME and analyzing in-situ data, we express our view that a filament identified in the COR FOV can be tracked further out in the heliosphere using \textit{J}-maps constructed from HI images. Our detailed examinations of remote and in-situ data reveal that F1 and F2 have been tracked as leading edge and filament in the COR2 FOV as such while Feature 1 and Feature 2 tracked as brightness enhancement (in the HI FOV) could be associated as sheath and filament material of the CME, respectively. We summarize that the features F1 and F2 observed in COR images are related to Feature 1 and Feature 2 in HI images, respectively.

Although, we believe that remotely observed features have been successfully associated with in-situ structures, it is important to remember the fundamental difference between both set of observations. There is a contribution to the imaging signal from all along the line of sight while the in-situ observations measure density at a particular distance and azimuth in the heliosphere at a time. Moreover, due to a fixed single point location of in-situ spacecraft, it is  difficult to claim that remotely tracked features are certainly intersected by the in-situ spacecraft unless we track them up to L1 which could not be done in the present case. From the present study, we also realize that the use of difference images for tracking of cavity (observed as magnetic cloud in in- situ) which is expected to lie between the two tracked features may be difficult. It is because; lack of density associated with cavity (flux rope) becomes indistinguishable from the brighter background on taking the running difference. However, from our study limited to a single CME, tracking of filaments seems to be possible using running difference images of SECCHI/COR and HI.

\begin{acknowledgements}
The authors thank T. A. Howard for his valuable comments and useful discussions. We thank the team members of \textit{STEREO}, \textit{ACE} and \textit{Wind} spacecraft for the data used in this paper. We acknowledge the UK Solar System Data Centre for providing the processed Level 2 \textit{STEREO}/HI data.  
\end{acknowledgements}



\begin{sidewaystable}
  \centering
{\scriptsize
 \begin{tabular}{p{3.0cm}|p{2.0cm}| p{3.0cm}| p{2.5cm}|p{2.5cm}|p{2.5cm}}
    \hline
		
 Method & Kinematics as inputs in DBM [t$_{0}$, R$_{0}$ (R$_\odot$), v$_{0}$ (km s$^{-1})$] &  Predicted arrival time using kinematics + DBM (UT) [$\gamma$ = 0.2 to 2.0 (10$^{-7}$ km$^{-1}$)] & Predicted transit speed at L1 (km s$^{-1}$)  [$\gamma$ = 0.2 to 2.0 (10$^{-7}$ km$^{-1}$)] & Error in predicted arrival time (hrs) [$\gamma$ = 0.2 to 2.0 (10$^{-7}$ km$^{-1}$)] &  Error in predicted transit speed (km s$^{-1}$)  [$\gamma$ = 0.2 to 2.0 (10$^{-7}$ km$^{-1}$)] \\  \hline

PP (\textit{STEREO-A})	& 10 Oct 03:30, 137, 230  & 12 Oct. 12:51 to 12 Oct. 03:22	 &  270 to 327   & 	23.7 to 14.2	 &  -85 to -28 \\ \hline

PP (\textit{STEREO-B})	& 09 Oct 12:40, 137, 360  & 11 Oct 05:14 to 11 Oct 05:21	 &  360 to 358   & 	-8 to -7.8	 &  5 to 3 \\ \hline
  
FP (\textit{STEREO-A})	& 09 Oct 10:30, 149, 380  &  10 Oct 18:54 to 10 Oct 19:27  & 	378 to 368  &	-18.2 to -17.6	& 23 to 13  \\ \hline

FP (\textit{STEREO-B})	& 08 Oct 23:39, 130, 464    & 10 Oct 11:05 to 10 Oct 15:28 & 	438 to 377 &	-26.3 to -21.7	& 83 to 22  \\ \hline
  
HM (\textit{STEREO-A})	& 09 Oct 11:30, 132, 285  & 11 Oct. 16:48 to 11 Oct. 12:43	 & 298 to 330  &	3.6 to -0.4	 &  -57 to -25  \\ \hline

HM (\textit{STEREO-B})	& 09 Oct 12:40, 158, 430  & 10 Oct 13:29 to 10 Oct 15:07  & 420 to 382 	 & -23.7 to -22.1	 &  65 to 27 \\  \hline
GT         & 09 Oct 12:40, 159, 450   & 10 Oct 12:03 to 10 Oct 14:06 & 436 to 385  &  -25.2 to -23.2  & 81 to 30   \\  \hline

TAS       &  09 Oct 12:40, 131, 280    & 11 Oct 19:21 to 11 Oct 14:41  & 295 to 330 &   6.1 to 1.4  & -60 to -25   \\  \hline
SSSE &  09 Oct 12:40, 154, 385 &  10 Oct 18:09 to 10 Oct 18:45	& 383 to 370 	& -19.1 to -18.5	 & 28 to 15    \\  \hline		

 \end{tabular}

\begin{tabular}{p{3.0cm}| p{3.0cm}| p{2.5cm}|p{2.5cm}|p{2.5cm}| p{2.0cm}}
\multicolumn{6}{c}{Time-elongation track fitting methods} \\  \hline
  
 Methods   & Best fit parameters [t$_{(\alpha = 0)}$, $\Phi$ ($^\circ$), v (km s$^{-1}$)]  &   Predicted arrival time at L1 (UT)  & Error in predicted arrival time  & Error in predicted speed at L1 (km s$^{-1}$) & Longitude ($^\circ$) 
   \\ \hline
	
	FPF (\textit{STEREO-A})   & 06 Oct 07:44, 90, 378 & 10 Oct 20:29 & -16.9  &  23      &  -7   \\  \hline
	FPF  (\textit{STEREO-B})  & 06 Oct 04:14, 56.2, 369 & 10 Oct 19:25  & -17.7  &    14   &  -21.5 \\  \hline
	 
	HMF (\textit{STEREO-A})   & 06 Oct 08:54, 125, 467 & 11 Oct 07:48 & -5.4  &    -18    &  -42   \\  \hline
	HMF  (\textit{STEREO-B})  & 06 Oct 05:51, 58, 378 & 11 Oct 01:35 &  -11.6  &    -10   &   -19.8 \\  \hline
	
	SSEF (\textit{STEREO-A})  & 06 Oct 08:40, 114, 436  & 11 Oct 09:02   & -4.2  &   -13  &  -31 \\  \hline
	SSEF (\textit{STEREO-B})  & 06 Oct 05:37, 57.4, 377 & 11 Oct 05:43  & -7.5  &  -12   &   -20.4   \\  \hline

\end{tabular}
}
\caption{\scriptsize{The kinematics input to the DBM and the resulting predicted arrival times and speeds (and errors therein) of Feature 2 at L1, corresponding to the extreme range of drag parameter. In the bottom panel, best fit parameters, predicted arrival times and speeds (and errors therein) estimated from time-elongation track fitting methods are given. The \textit{STEREO-A} and \textit{B} shown in parentheses for each method denotes the spacecraft from which data is   used. Negative (positive) errors in predicted arrival time correspond to a predicted arrival time that is before (after) the actual CME arrival time determined from in-situ  measurements. Negative (positive) errors in predicted  speed correspond to a predicted speed that is less that (more than) than actual CME  speed at L1.}}
\label{KinarrL1}
\end{sidewaystable}

\newpage

\end{document}